\documentclass[a4paper,11pt]{article}
\usepackage{graphicx}
\usepackage{algorithm}
\usepackage{algorithmicx}
\usepackage{algpseudocode}
\usepackage{amsmath}
\usepackage{epstopdf}
\usepackage{times}

\title{Count-Min-Log sketch: Approximately counting with approximate counters}
\author{Guillaume Pitel, eXenSa\\
		\texttt{guillaume.pitel@exensa.com} 
		\and 
		Geoffroy Fouquier, eXenSa\\
		\texttt{geoffroy.fouquier@exensa.com}}

\begin{document}

\maketitle

\begin{abstract}
Count-Min Sketch \cite{cormode2005improved} is a widely adopted algorithm for approximate event counting in large scale processing. However, the original version of the Count-Min-Sketch (CMS) suffers of some deficiences, especially if one is interested by the low-frequency items, as in text-mining related tasks. Several variants of CMS \cite{goyal2012sketch} have been proposed to compensate for the high relative error for low-frequency events, but the proposed solutions tend to correct the errors instead of preventing them. In this paper, we propose the Count-Min-Log sketch, which uses logarithm-based, approximate counters \cite{morris1978counting,flajolet1985approximate} instead of linear counters to improve the average relative error of CMS (with conservative update) at constant memory footprint. 
\end{abstract}

{\let\thefootnote\relax\footnote{Copyright \copyright 2015 eXenSa. All rights reserved.}}
\section{Introduction}

Count-Min Sketch \cite{cormode2005improved} (CMS) is a widely adopted algorithm for approximate event counting in large scale processing. With proved bounds in terms of mean absolute error and confidence, one can easily design a constant size sketch as an alternative to expensive exact counting for a setting where the total number of event types is approximately known.

CMS is used in many applications, often with a focus on high frequency events \cite{cormode2005s}. However, in the domain of text-mining, highest frequency events are often of low interest: frequent words are often grammatical, highly polysemous or without any interesting semantics, while low-frequency words are more relevant. 

As a matter of fact, a common regularizations in text-mining consist in computing the TF-IDF \cite{salton1983introduction} (equation \ref{eqn:TFIDF}) for term/document relevance, Pointwise Mutual Information \cite{xu2007study} (equation \ref{eqn:PMI}) or Log-likelihood Ratio \cite{dunning1993accurate} between two words in order to estimate the importance of their cooccurrence.

\begin{subequations}
 	\label{eqn:TFIDF}
 	\begin{align}
 	idf_i = \log \frac{|D|}{|d_j: t_i \in d_j|} \\
 	tfidf_{i,j} = tf_{i,j} \cdot idf_i
 	\end{align}
 \end{subequations}
 
 \begin{subequations}
 	\label{eqn:PMI}
 	\begin{align}
 	pmi_i,j = \log \frac{p(i,j)}{p(i)p(j)}
 	\end{align}
 \end{subequations}

In TF-IDF, $|D|$ is the number of documents in the corpus, $tf_{i,j}$ is the frequency of term $t_i$ in document $d_j$ and $|d_j: t_i \in d_j|$ is the number of documents containing the term $t_i$. In PMI, $p(i,j)$ is the probability that words $i$ and $j$ appear in a cooccurrence window and $p(i)$ the probability to find $i$ in the corpus.

Those formulas show that higher frequency words will induce a relatively lower value at the end. Moreover, they all use a logarithm, which shows that only the order of magnitude is important.

The original version of the Count-Min-Sketch (CMS) suffers from some deficiences, especially when one is interested not by the high frequency events, but by the low frequency ones, such as these tasks above. A variant of CMS \cite{goyal2012sketch} has been proposed to compensate for the high relative error for low-frequency events, but the solutions explored tend to correct the errors instead of preventing them. 

\section{Count-Min-Log Sketch with Conservative Update}

We propose a variant of the Count-Min Sketch with conservative update that uses logarithm-based, approximate counters \cite{morris1978counting,flajolet1985approximate} instead of linear counters to improve the average relative error of CMS at constant memory footprint. The principle is to use exactly the same structure than Count-Min Sketch with conservative update, replacing only the classical binary counting cells by log counting cells. With this modification, the \textsc{Update} and \textsc{Query} procedures becomes respectively algorithm \ref{alg:CMLS-UPDATE} and algorithm \ref{alg:CMLS-QUERY}.

\begin{algorithm}
	\caption{Count-Min-Log Sketch UPDATE}\label{alg:CMLS-UPDATE}
	\renewcommand{\algorithmicrequire}{\textbf{Input:}}
	\renewcommand{\algorithmicensure}{\textbf{Output:}}
	\begin{algorithmic}[1]
		\Require sketch width $w$, sketch depth $d$ , log base $b > 1$, independent hash functions $h_{1..d}:U \rightarrow \{1 \ldots w\}$  
		\Function{IncreaseDecision}{$c$}
			\State \Return True with probability $b^{-c}$, else False 
		\EndFunction
		\Function{Update}{$e$}
		\State $c \gets \min_{1 \le k \le d}{sk[k,h_k(e)]}$
		\If{\Call{IncreaseDecision}{c}}
	    \For{$k \gets 1 \ldots d$} 
	    \If{$sk[k,h_k(e)] = c$} 
	    \State $sk[k,h_k(e)] \gets c+1$
	    \EndIf
	    \EndFor
		\EndIf
		\EndFunction
		
	\end{algorithmic}
\end{algorithm}

\begin{algorithm}
	\caption{Count-Min-Log Sketch QUERY}\label{alg:CMLS-QUERY}
	\renewcommand{\algorithmicrequire}{\textbf{Input:}}
	\renewcommand{\algorithmicensure}{\textbf{Output:}}
	\begin{algorithmic}[1]
		\Require sketch width $w$, sketch depth $d$ , log base $b > 1$, independent hash functions $h_{1..d}:U \rightarrow \{1 \ldots w\}$  
		\Function{PointValue}{$c$}
		\If{$c = 0$}
		\State \Return $0$
		\Else
		\State \Return $b^{c-1}$
		\EndIf
		\EndFunction
		\Function{Value}{$c$}
		\If{$c \le 1$}
		\State \Return \Call{PointValue}{$c$}
		\Else
		\State $v \gets $\Call{PointValue}{$c+1$}
		\State \Return $\frac{1 - v}{1-x}$
		\EndIf		
		\EndFunction
		
		\Function{Query}{$e$}
		\State $c \gets \min_{1 \le k \le d}{sk[k,h_k(e)]}$
		\State \Return \Call{Value}{$c$}
		\EndFunction
	\end{algorithmic}
\end{algorithm}

The rationale behind this variant lies as follows: 
\begin{enumerate}
	\item The original CMS uses as many bits (usually 32 or 64) to represent low values than high values. However, in skewed distributions like those of natural languages, low values are much more frequent than high values, and they use less bits. As a consequence, one can consider using smaller counters, and thus increasing the number of counters for the same storage space.
	\item Furthermore, and this is especially the case for highly skewed distributions like the ones of Zipfian data, Count-Min Sketch estimates high frequency events very well (Count-Min Sketch with Conservative Update is even better in this regard), but very poorly low-frequency events. As a consequence, estimates of Pointwise Mutual Information, for instance, are largely off for low frequency items, which may cause several problems for large scale NLP tasks.
\end{enumerate} 

These observations led to a first conclusion, shared by \cite{goyal2012sketch}, which is that for NLP tasks, the error metric that should be used on Point Query estimations is the Average Relative Error (ARE), not the Root Mean Square Error (RMSE). This led in turn to the following hypothesis: replacing the linear counters by logarithmic counters should incur, under some conditions, an improvement over the ARE.

Another way to evaluate the precision of an approximate counting sketch for NLP task consists in directly measuring the error on the Pointwise Mutual Information.

\section{Empirical evaluation}
\subsection{Data}
We have verified this hypothesis empirically in the following setting: we count unigrams and bigrams of 500k words of the 20newsgroups corpus \cite{Lang95}. The small corpus analyzed contains 233k counted elements, composed of 183k bigrams and 50k unigrams. 

In the following, the ideal perfect count storage size corresponds, for a given number of elements, at the minimal amount of memory to store them perfectly, in a ideal settings. A high-pressure setting corresponds to a setting where the memory footprint is lower than the ideal perfect count storage size for the same number of elements.

\subsection{Variants}
We compare the estimates of three sketches: 
\begin{description}
\item[CMS-CU] is the classical linear Count-min Sketch with Conservative Update, 
\item[CMLS16-CU] is the Count-min-log Sketch with Conservative Update using a logarithmic base of 1.00025 and 16bits counters, and 
\item[CMLS8-CU] is the Count-min-log Sketch with Conservative Update using a logarithmic base of 1.08 and 8bits counters.
\end{description}

\subsection{Error on counts}

\begin{figure*}
\centering
\includegraphics[width=0.8\linewidth]{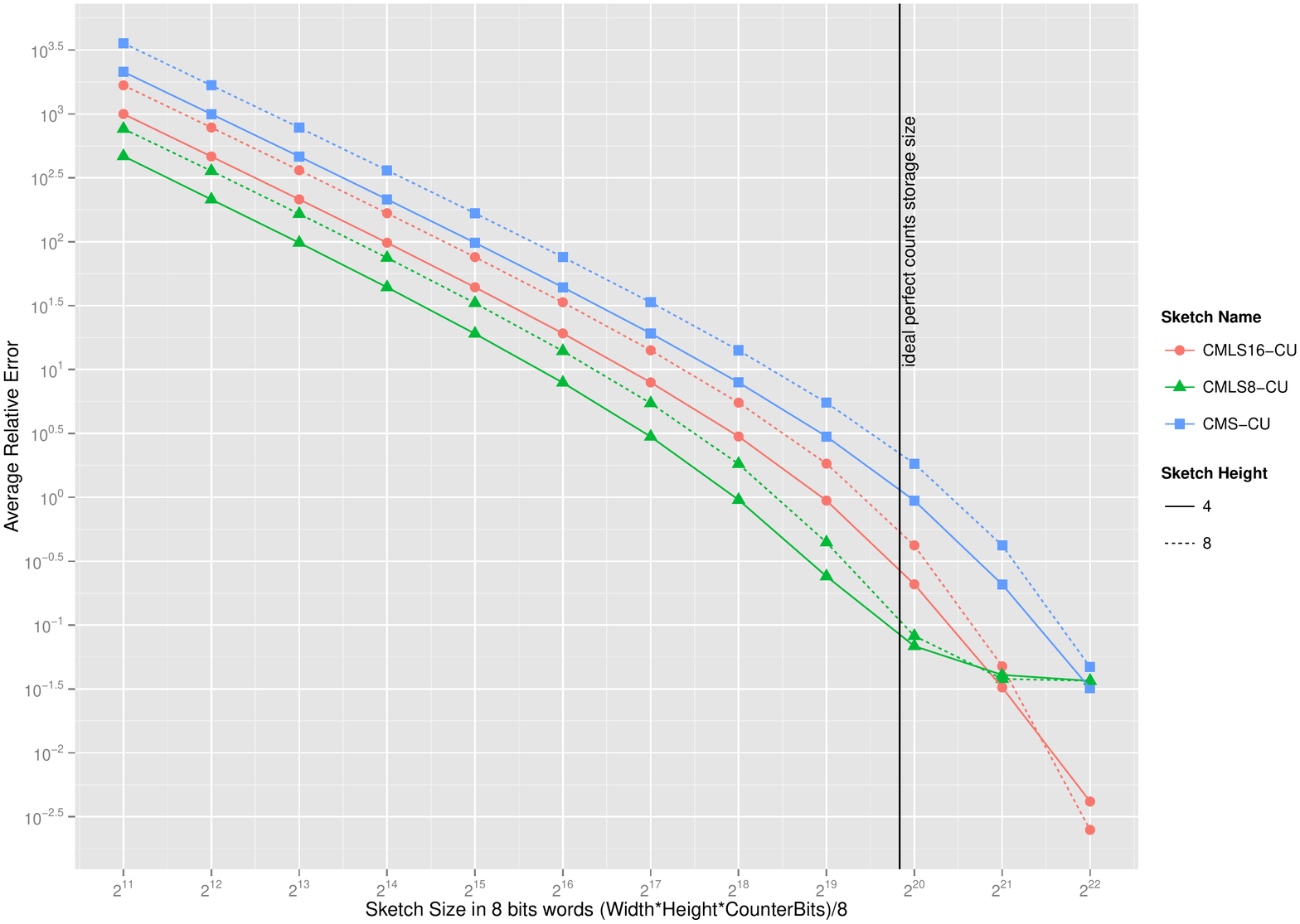}
\caption[Average Relative Error of sketches counts]{Average Relative Error of estimated counts with Count-Min Sketch (CMS-CU, blue lines), Count-Min-Log 16bits (CMLS16-CU, red lines) and Count-Min-Log 8bits (CMLS16-CU, green lines). The bold vertical line corresponds to the ideal perfect counts storage size.}
\label{fig:Expe-ARE-MEAN}
\end{figure*}

The results for the Average Relative Error on simple counts are shown on figure \ref{fig:Expe-ARE-MEAN}. The vertical line indicates the storage needed to memorize perfectly all the counts (the extra memory required for accessing the counters is not taken into account).

This experiments show that before the perfect storage mark, the estimate error of the CMLS16-CU is approximately 2 to 4 times lower than the error of the estimate of the CMS-CU. The CMLS8-CU error improvement over CMS-CU is in the range of 7 to 12 times, however, the CMLS8-CU reaches a minimal ARE of $10^{-1.5}$ and stops improving, due to the residual error caused by approximate counting. 

\subsection{Error on PMI}

\begin{figure*}
\centering
\includegraphics[width=0.8\linewidth]{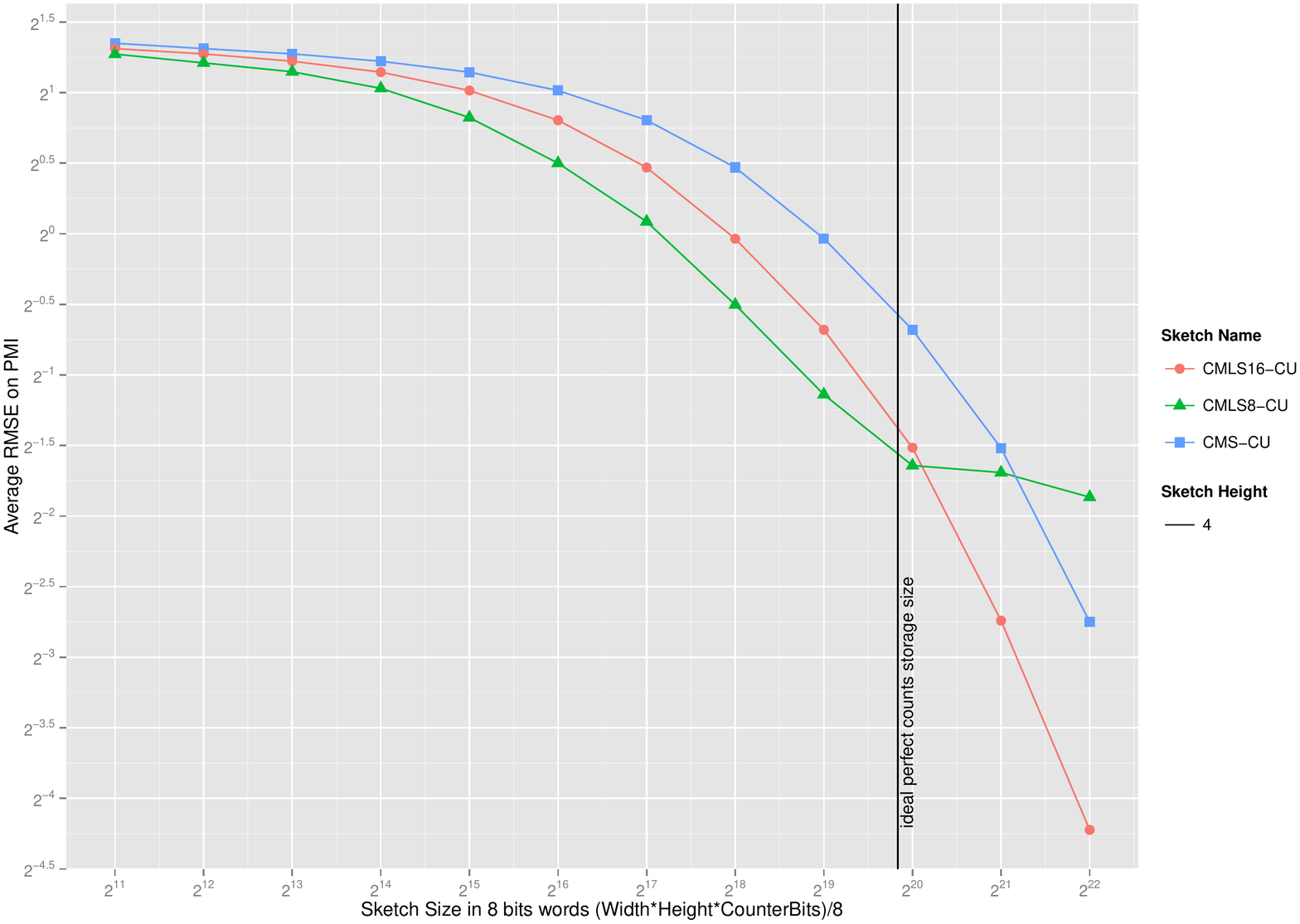}
\caption[RMSE of estimated PMI versus exact PMI]{Root Mean Square Error of estimated PMI with Count-Min Sketch (CMS-CU, blue line), Count-Min-Log 16bits (CMLS16-CU, red line) and Count-Min-Log 8bits (CMLS8-CU, green line). The bold vertical line corresponds to the ideal perfect counts storage size.}
\label{fig:Expe-RMSE}
\end{figure*}

In a second step, we compute the Pointwise Mutual Information of the bigrams, and the error between the estimated PMI using counts from the sketch versus using the exact counts. The results for RMSE on estimated PMI are illustrated in figure \ref{fig:Expe-RMSE}.

These results show that, with sketches near or smaller than the theoretical size of a perfect storage, CMLS16-CU (respectively CMLS8-CU) outperforms CMS-CU by a factor of about 4 (resp. 10) on the RMSE of the PMI.

\begin{figure*}
\centering
\includegraphics[width=0.7\linewidth]{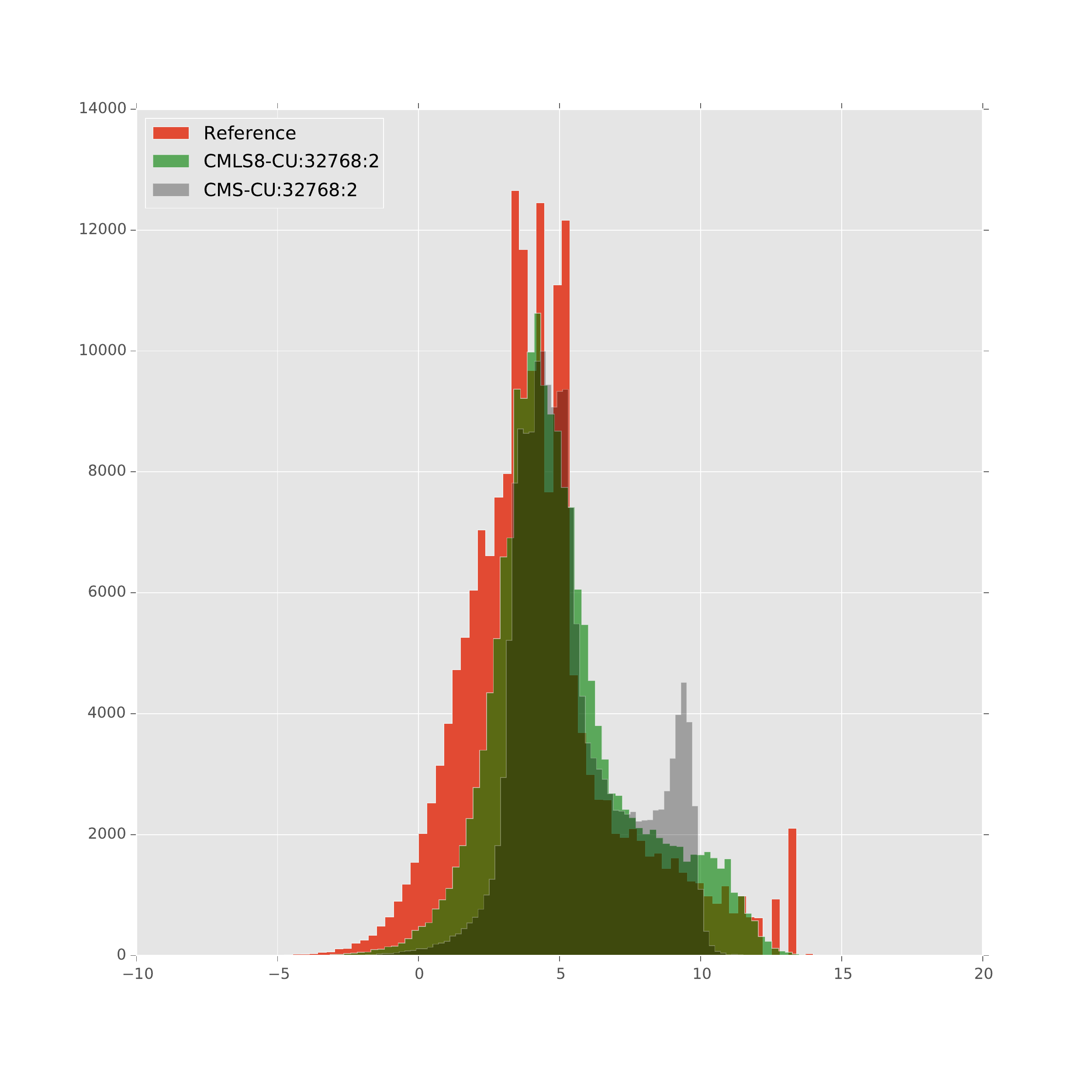}
\caption[Histogram of PMI values]{Histograms of PMI values estimated for Count-Min Sketch (CMS-CU, grey area) and Count-Min-Log 8bits (CMLS8-CU, green area) sketches with 32kb storage, 2 levels. The area in red is the reference PMI computed with exact counts.}
\label{fig:histPmi}
\end{figure*}

The histograms of PMI values for each sketch are illustrated by figure \ref{fig:histPmi}, showing that for equivalent storage, the CMS-CU presents a very distorted histogram on the right part (the interesting part for NLP tasks), while the CMLS8-CU is much closer to the reference.

\section{Conclusion and perspectives}

We have proposed a simple variant of the classical Count-Min-Sketch with Conservative Update which present significant improvement for counts Average Relative Error and RMSE on the Pointwise Mutual Information. While the gain is always clear for high pressure setups, when the available storage is less or equal the ideal storage size, the residual error due to the approximate counting is an absolute lower bound of the error, which can be hit at different sketch sizes depending on the logarithm base.

The next steps of this work are the following:
\begin{itemize}
\item Compare results of PMI estimations only for interesting values (i.e. over a given threshold), since we have observed in figure \ref{fig:histPmi} that CMS-CU seems to be particularly far from the reference on the right side of the histogram.
\item Evaluate the speed difference of our variant compared to CMS-CU.
\end{itemize}

Additionally, we are investigating two other directions: 
\begin{enumerate}
\item Hierarchical storage cells with more cells to store least significant bits and less cells for most significants bits.
\item Probabilistic update rule: we have observed that the ratio between smallest and second smallest estimates is correlated with the error. We want to try a probabilistic update rule that will take this ratio into account.
\end{enumerate}

\bibliographystyle{plain}
\bibliography{article-arxiv}
\end{document}